\documentclass[prd,aps,floats,twocolumn,prd,showpacs,nofootinbib]{revtex4}

\usepackage{amssymb}
\usepackage{amsmath}

\usepackage{graphicx}
\usepackage{graphics}
\usepackage{dcolumn}
\usepackage{color}
\usepackage{rotate}
\usepackage{fancyhdr} 
\usepackage{graphicx}

\newcommand{\e}{\epsilon}
\newcommand{\M}{M_{\mathrm{pl}}}
\newcommand{\La}{\Lambda}

\begin{document}

\title{Equation-of-State Parameter for Reheating}

\author{Julian B. Mu\~noz\footnote{julianmunoz@jhu.edu} and Marc Kamionkowski\footnote{kamion@jhu.edu}}
\affiliation{Department of Physics and Astronomy, Johns
     Hopkins University, 3400 N.\ Charles St., Baltimore, MD 21218}

\date{\today}

\begin{abstract}
Constraints to the parameters of inflation models are often
derived assuming some plausible range for the number---e.g., $N_k=46$ to
$N_k=60$---of $e$-folds of inflation that occurred
between the time that our current observable Universe exited the
horizon and the end of inflation.  However, that number is, for
any specific inflaton potential, related to an effective
equation-of-state parameter $w_{\mathrm{re}}$ and temperature $T_{\mathrm{re}}$, for reheating.  Although the
physics of reheating is highly uncertain, there is a finite
range of reasonable values for $w_{\mathrm{re}}$.  Here we show that
by restricting $w_{\mathrm{re}}$ to this range, more stringent constraints
to inflation-model parameters can be derived than those obtained
from the usual procedure.  To do so, we focus in this work in
particular on natural inflation and inflation with a Higgs-like
potential, and on power law models as limiting cases of those.  As one example, we show that the lower limit to the
tensor-to-scalar ratio $r$, derived from current measurements of
the scalar spectral index, is about 20\%-25\% higher (depending on the model)
 with this procedure than with the usual approach.
\end{abstract}


\maketitle

\section{Introduction}

Models of inflation that rely on the slow rolling of a single
scalar field have become the canonical family of models for
inflation \cite{Linde:1983gd,0811.1989,1404.1065}.  These models are specified by a potential-energy
density $V(\phi)$ given as a function of the inflaton field
$\phi$.  As long as the slow-roll conditions, which require that
the slope and curvature of $V(\phi)$ are sufficiently small, are satisfied, the
Universe inflates.  Inflation then ends and is followed by a
period of reheating (see Ref.~\cite{Allahverdi:2010xz} for a
review) that converts the energy density in the inflaton to the
thermal bath, at a reheating temperature $T_{\mathrm{re}}$, that fills the
Universe at the beginning of the standard radiation-dominated
epoch.

In the canonical reheating scenario \cite{elementary-theory},
oscillations of the inflaton around the minimum of its potential
correspond to massive inflaton particles, and these particles
then decay to the plasma of Standard Model particles that
compose the radiation-dominated Universe.  However, the physics
of reheating may be far more complicated.  For example,
different rates for different types of decays into different
Standard Model particles may yield different clocks for starting
the usual radiation-dominated epoch.  There may be a preheating
stage \cite{1410.3808}, where there is a resonant
production of particles \cite{Kofman:1994rk}, which can enhance
the inflaton decay via scattering \cite{hep-ph/9704452}, or
where inhomogeneous modes may be excited \cite{hep-ph/9705357}.
Turbulence may also play a role \cite{turbulent-thermalization}.
It is generally assumed that the reheat temperature is above the
electroweak transition (presumably so that weak-scale dark
matter can be produced).  More conservatively, though, the
reheat temperature must be above an MeV, the temperature of
big-bang nucleosynthesis, the earliest time for which we have
clear empirical relics.  The theoretical uncertainty in
reheating is often taken into account, in the 
consideration of experimental constraints to inflation models, 
by surmising some reasonable range---e.g., $N_k=46$ to
$N_k=60$---for the number $N_k$ of $e$-folds of inflation between
the time that our observable horizon exited the horizon during
inflation and the end of inflation.  The upper limit to this
range arises if inflaton oscillations reheat the Universe
instantaneously to a grand unified theory-scale temperature, and the lower limit arises 
if reheating is closer to the electroweak scale.

Here we consider an alternative approach where we parametrize
the cosmic fluid during reheating by an effective
equation-of-state parameter $w_{\mathrm{re}}$, that tells us how its energy
density ($\rho\propto a^{-3(1+w_{\mathrm{re}})}$) decays during this epoch.
In the canonical-reheating scenario $w_{\mathrm{re}}=0$, but numerical
studies of thermalization indicate a possibly broader range of
values $0\lesssim w_{\mathrm{re}}\lesssim 0.25$ \cite{hep-ph/0507096}.
By demanding that the equation-of-state parameter fall within
this range, we infer slightly better constraints to inflation
models than in the usual approach wherein some overly permissive
range of $N_k$ is assumed.  The approach we use here was discussed
in
Refs.~\cite{Dodelson:2003vq,Martin:2010kz,Adshead:2010mc,Mielczarek:2010ag,Easther:2011yq}
and applied post-Planck to power-law potentials in
Ref.~\cite{Dai:2014jja}.  
In this paper we explore this approach and show its general validity for single
field inflation models. As an example, we apply it to
study constraints to the parameter space for natural inflation
\cite{Adams:1992bn,Freese:2014nla} and Higgs-like inflation models \cite{Kaplan:2003aj}.  We show in particular that the
lower limit to the tensor-to-scalar ratio $r$ inferred from
current measurements of $n_s$ should be a bit higher (by about
$25\%$) if we restrict the value of $w_{\mathrm{re}}$ to the range
suggested by reheating theory.

The structure of this paper is as follows.  In Section II we discuss 
how the effective reheating equation-of-state parameter imposes 
restrictions to the model.  In Section III we review the natural and 
Higgs-like inflaton potentials we focus upon in this paper.  
Section IV presents the results, and in Section V we make concluding remarks.

\section{Reheating}

\begin{figure}[htbp]
\centering
\hspace{-30pt}
\includegraphics[width=1.1\linewidth]{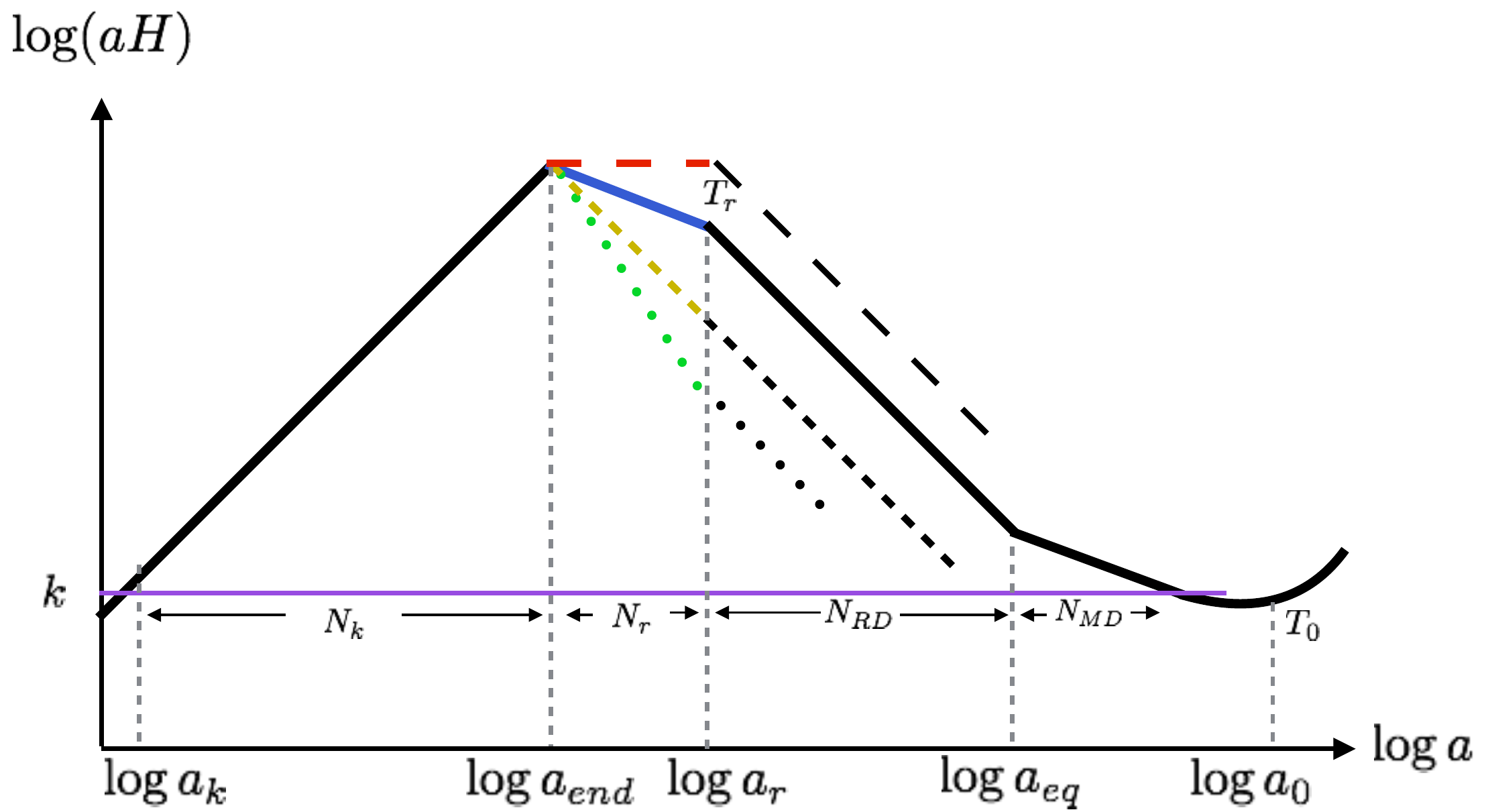}
 \caption
   {Comoving Hubble parameter $aH$ versus scale factor $\log a$.  
   A comoving mode with wavenumber $k$ exits the horizon during 
   inflation when $k=aH$ and then reenters during matter domination.  
   Different equations of state for reheating are plotted: canonical reheating 
   ($w_{\mathrm{re}}=0)$ in blue (solid); $w_{\mathrm{re}}=-1/3$ in red (long dash); $w_{\mathrm{re}}=1/3$ in brown 
   (short dash); and the limiting case $w_{\mathrm{re}}=1$ in green (dotted).} 
  \label{fig:Horizons}
\end{figure}

Fig.~\ref{fig:Horizons} shows the comoving Hubble parameter $aH$
with time \cite{Baumann:2009ds}.  It grows for $N_k$ $e$-folds
during inflation with a time dependence that is fixed given a
specific inflaton potential $V(\phi)$.  It then decreases for
$N_{\rm re}$ $e$-folds of expansion during which the energy
in the inflaton potential is dissipated into a radiation bath.
The standard radiation-dominated era then proceeds for $N_{\rm
RD}$ $e$-folds before the advent of matter domination (and then
dark-energy domination).  It is clear from the Figure that the
number of $e$-folds of expansion between the time that a given
scale exits the horizon and the end of inflation is related to
the number of $e$-folds since the end of inflation until that
scale re-enters the horizon during matter/radiation-domination.
The expansion history also determines the evolution of the
energy density, and a second relation can be obtained from a
given expansion history by demanding the proper relation between
the energy density during inflation and the energy density today.

A consistent model for inflation must have an inflaton potential
$V(\phi)$ that at some point steepens so that the slow-roll
condition $\epsilon<1$ (where $\epsilon=(V'/V)^2/2\M^2$ is the
slow-roll parameter and $\M$ is the reduced Planck mass) breaks down, at which point inflation ends.
The number of $e$-folds between the time that a comoving scale
$k$ exits the horizon and the end of inflation is
\begin{equation}
     N_k = \int_{\phi_k}^{\phi_{\rm end}} \, \frac{H\,
     d\phi}{\dot\phi},
\end{equation}
where $\phi_k$ is the inflaton value when $k$ exits the horizon,
$H(\phi)$ is the Hubble parameter, and the dot denotes a
derivative with respect to time $t$.  The Hubble parameter can
then be written in terms of the inflaton potential using the Friedmann
equation, $H^2\simeq V/(3M_{\rm pl}^2)$, and $\dot\phi$ is
evaluated through the slow-roll equation, $3 H
\dot\phi+V'(\phi)\simeq0$, where the prime denotes derivative with
respect to $\phi$.  The values of the scalar spectral index
$n_s$ and tensor-to-scalar ratio $r$ can be obtained as a
function of $N_k$.  Given the relation between $N_k$ and the
number of post-inflation $e$-folds of expansion, the value of
$N_k$ relevant for cosmic microwave background measurements is a fixed function of $n_s$ once a given
reheating history (specified by $w_{\mathrm{re}}$ and the reheat temperature
$T_{\mathrm{re}}$) is assumed. Below we will use the fairly well-determined value
of $n_s$ to infer, for a given reheat scenario, the
inflaton-potential parameters and from them the allowable values
of $r$.

Let us consider the pivot scale $k=0.05~{\rm Mpc}^{-1}$ at which
Planck determines $n_s$ \cite{Ade:2013uln}.  The comoving Hubble scale $a_k H_k =
k$ when this mode exited the horizon is related to that, $a_0
H_0$, of the present time by,
\begin{equation}
     \frac{k}{a_0 H_0} = \frac{a_k}{a_{\rm end}} \frac{a_{\rm
     end}}{a_{\rm re}} \frac{a_{\rm re}}{a_{\rm eq}}
     \frac{a_{\rm eq} H_{\rm eq}}{a_0 H_0} \frac{H_k}{H_{\rm
     eq}},
\end{equation}
where quantities with subscript $k$ are evaluated at horizon
exit. The other subscripts refer to the end of inflation (${\rm
end}$), reheating (${\rm re}$), radiation-matter equality (${\rm
eq}$), and the present time ($0$). Using $e^{N_k}=a_{\rm end}/a_k$,
$e^{N_{\rm re}}=a_{\rm re}/a_{\rm end}$ and $e^{N_{\rm
RD}}=a_{\rm eq}/a_{\rm re}$, we obtain the constraint,
\begin{equation}
\label{eq:efolds-eq-1}
   \ln\frac{k}{a_0 H_0} = - N_k - N_{\rm re} - N_{\rm RD} +
   \ln\frac{a_{\rm eq} H_{\rm eq}}{a_0 H_0} + \ln
   \frac{H_k}{H_{\rm eq}},
\end{equation}
on the total expansion~\cite{Liddle:2003as}.
The Hubble parameter during inflation is given by $H_k = \pi
\M \left(r A_s\right)^{1/2}/\sqrt 2$, with the primordial
scalar amplitude $\ln(10^{10} A_s) = 3.089^{+0.024}_{-0.027}$
from Planck~\cite{Ade:2013uln}.

The energy density $\rho_{\mathrm{end}}$ at the end of inflation
is related to the energy density $\rho_{\mathrm{re}}$ at the end of
reheating by the equation-of-state parameter $w_{\mathrm{re}}$ during reheating via
\begin{equation}
     \dfrac{\rho_{\mathrm{re}}} {\rho_{\mathrm{end}}}= \exp[-3 N_{\rm re} (1+w_{\mathrm{re}}) ],
\label{eq:energ}
\end{equation}
where $N_{\rm re}$ is the number of $e$-folds of expansion
during reheating.  

The energy density at the end of inflation is obtained from
\begin{equation}
     \rho_{\mathrm{end}}=(1+\lambda)V_{\mathrm{end}},
\end{equation}
where the ratio $\lambda$ of kinetic to potential energies at
the end of inflation is
\begin{equation}
     \lambda= \dfrac 1 {3/\e-1}.
\end{equation}
When inflation ends ($\e\approx1$), we have $\lambda \approx 1/2$.

We next calculate the energy density at reheating.  Assuming
conservation of entropy,
\begin{equation}
    g_{\mathrm{s,re}} T_{\mathrm{re}}^3 = \left( \dfrac{a_0}{a_{\mathrm{re}}}\right)^3 \left(2 T_0^3 +
    \dfrac {21} 4 T_{\nu,0}^3\right),
\end{equation}
where $g_{\mathrm{s,re}}$ is the effective number of relativistic degrees
of freedom at reheating, and $T_{\nu,0} = (4/11)^{1/3}T_0$ is
the current neutrino temperature.  Thus,
\begin{equation}
     \dfrac{T_{\mathrm{re}}}{T_0} = \left(
     \dfrac{43}{11g_{\mathrm{s,re}}}\right)^{1/3}
     \dfrac{a_0}{a_{\mathrm{eq}}}\dfrac{a_{\mathrm{eq}}}{a_{\mathrm{re}}}.
\label{eq:tem}
\end{equation}
Since the energy density at reheating is $\rho_{\mathrm{re}} = (\pi^2
g_{\mathrm{re}}/30) T_{\mathrm{re}}^4$, we plug Eq.~(\ref{eq:tem}) into
Eq.~(\ref{eq:energ}) to get the number $N_{\rm re}$ of $e$-folds
during reheating as a function of the number $N_{\rm RD}$ of
$e$-folds during radiation domination. Plugging that into
Eq.~(\ref{eq:efolds-eq-1}) we obtain finally,
\begin{eqnarray}
     N_{\rm re} &=& \dfrac 4{1-3\,w_{\mathrm{re}}} \left [ - N_k - \log(\dfrac k
     {a_0 T_0}) -\dfrac 1 4
     \log\left(\dfrac{30}{g_{\mathrm{re}}\pi^2}\right) \right. \nonumber
     \\
     & & \left. -\dfrac 1 3 \log\left(\dfrac{11 g_{\mathrm{s,re}}}{43}\right) - \dfrac 1 4 \log\left(V_{\mathrm{end}}\right)-
     \right. \nonumber \\
      & & \left. - \dfrac 1 4
      \log(1+\lambda) + \dfrac 1 2 \log\left(\dfrac{\pi^2 r A_s}{2}\right)
      \right ], 
\label{eq:nr}
\end{eqnarray}
where $g_{\mathrm{re}}$ and $g_{\mathrm{s,re}}$ can be both taken to be $\approx 100$ and we will use
$k=0.05\,\mbox{Mpc}^{-1}$ throughout the paper, albeit keeping the subindex $k$ in $N_k$ to avoid confusion.
Then using Eq.~(\ref{eq:energ}), the
reheating temperature is, 
\begin{equation}
     T_{\mathrm{re}} = \exp\left[{-\frac 3 4 (1+w_{\mathrm{re}})N_{\rm re}}\right]\left(\dfrac
     3{10\pi^2}\right)^{1/4} (1+\lambda)^{1/4}
     V_{\mathrm{end}}^{1/4}.
\end{equation}

\begin{figure*}[htbp!]
\centering
\hspace{-0.5cm}
\includegraphics[width=0.3\linewidth]{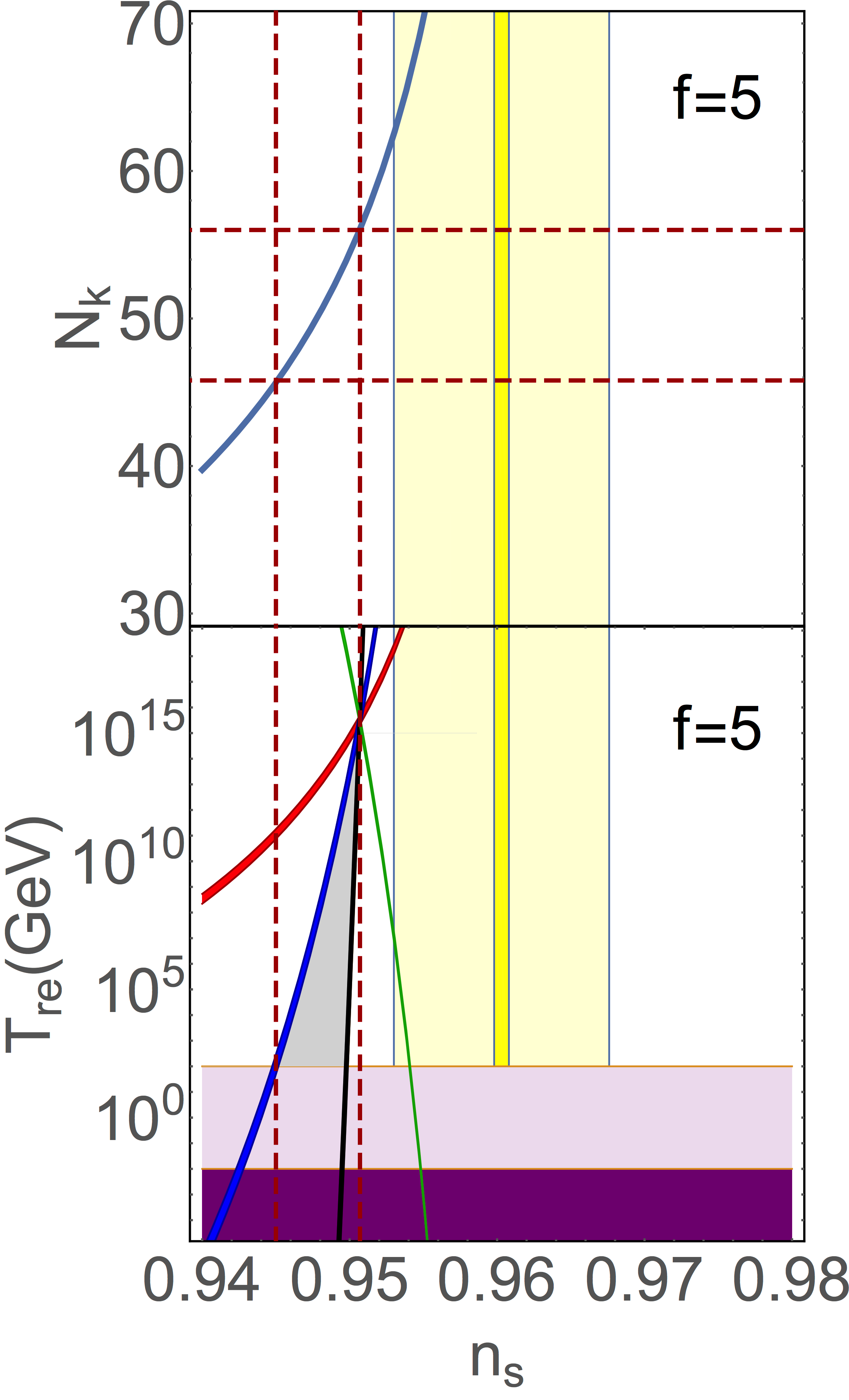}
\hspace{0.5cm}
\includegraphics[width=0.3\linewidth]{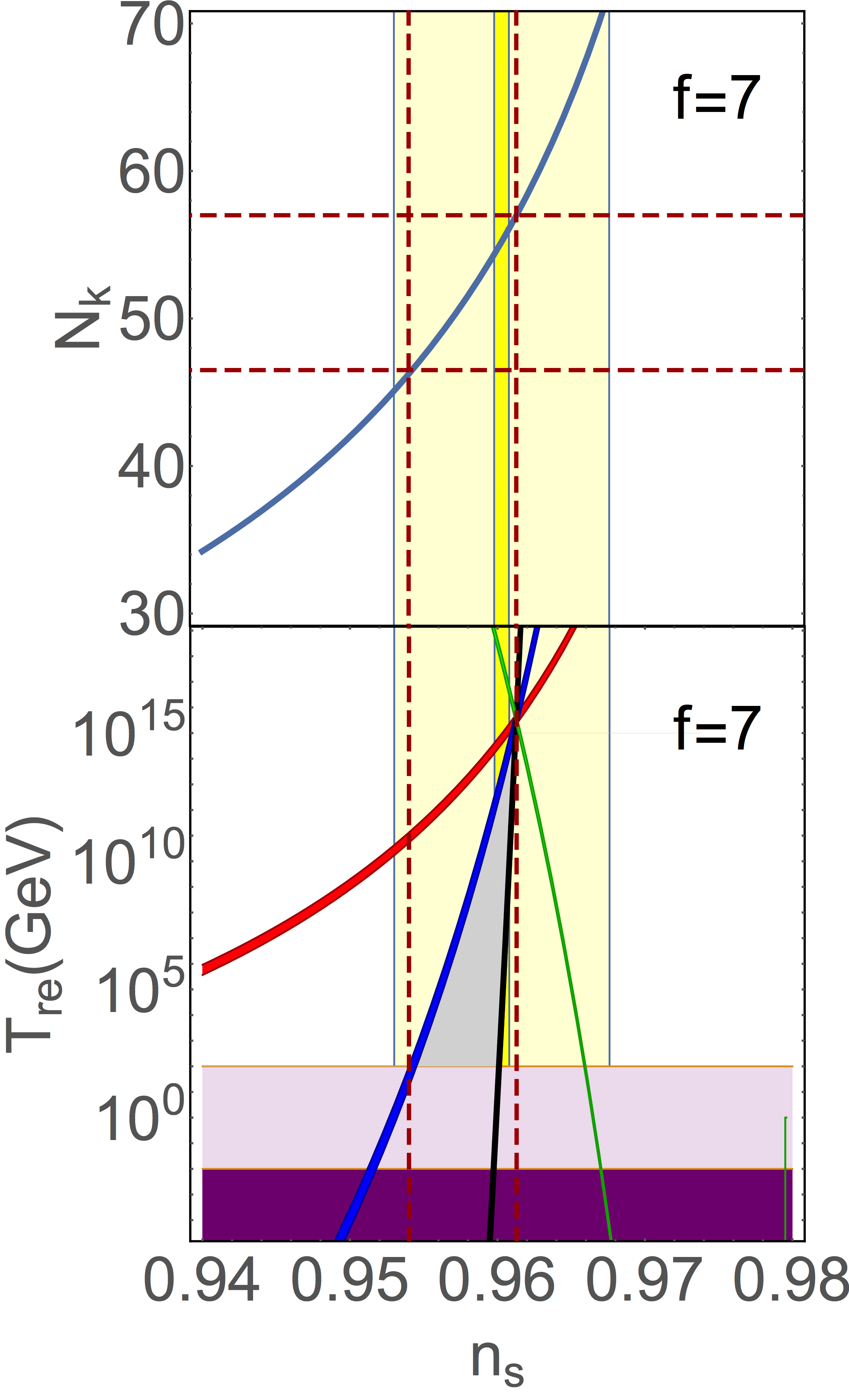}
\hspace{0.5cm}
\includegraphics[width=0.3\linewidth]{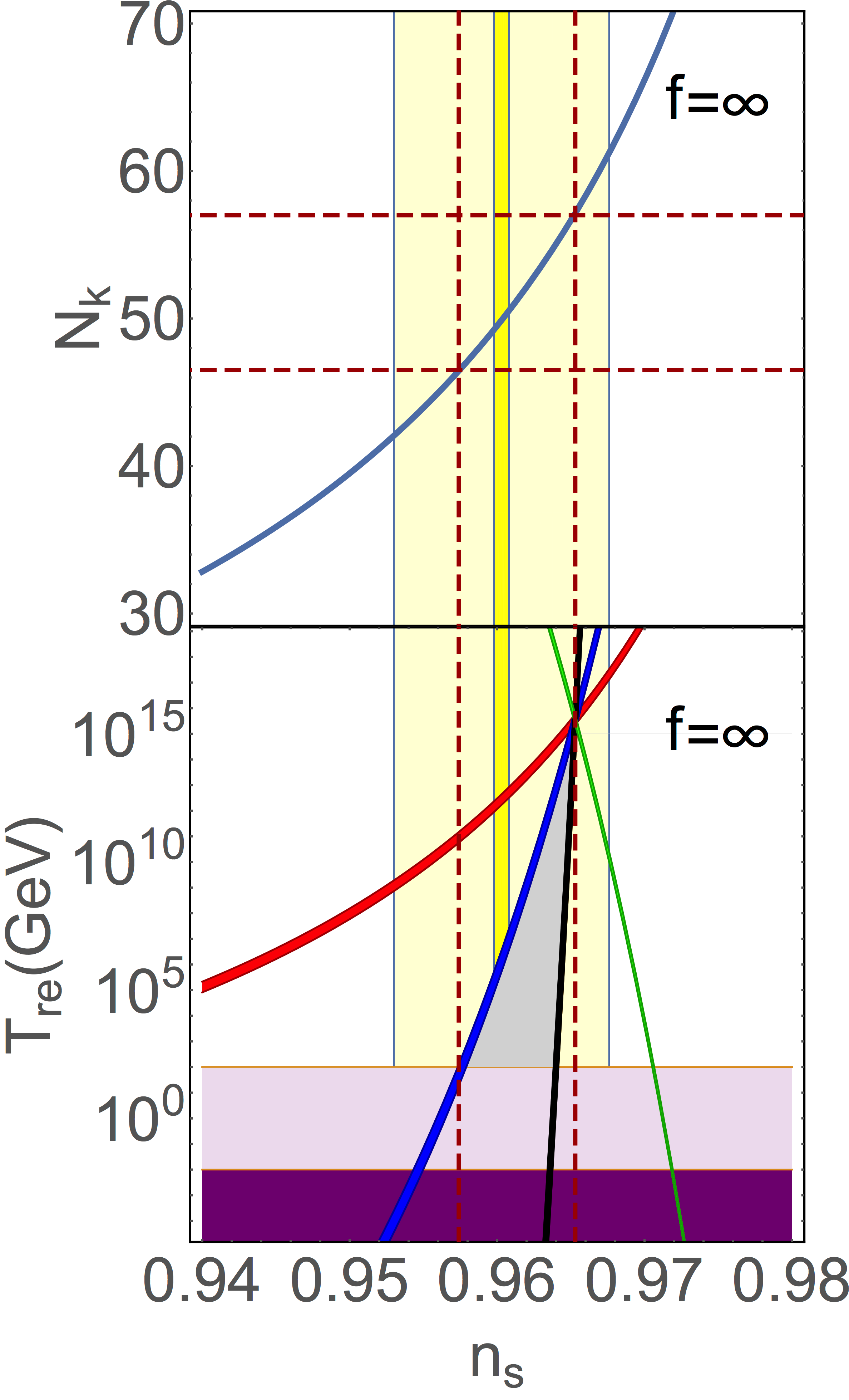}
\caption{In the lower panels we plot the reheat temperature
     $T_{\rm re}$ for natural inflation as 
     determined by matching the number of $e$-folds during and
     after inflation. Results are shown for decay constants
     $f=5\,\M$, 7 $\M$, and $\infty$, where the latter corresponds to the $m^2\phi^2$ limit. 
     Four different effective
     equation-of-state parameters $w_{\mathrm{re}}$ for reheating are
     considered in each case: from left to right in their
     intersection with the bottom of the plots they are $w_{\mathrm{re}}
     =-1/3$ (red), $w_{\mathrm{re}}=0$ (blue), $w_{\mathrm{re}}=0.25$
     (black), and $w_{\mathrm{re}}=1$ (green).  The values $w_{\mathrm{re}}=-1/3$ and $w_{\mathrm{re}}=1$
     bracket the very most conservative allowed range of values
     for $w_{\mathrm{re}}$, while $w_{\mathrm{re}}=0$ and $w_{\mathrm{re}}=0.25$ bracket the range suggested by the literature
     on reheating.  All curves intersect at the point where
     reheating occurs instantaneously, and the $w_{\mathrm{re}}=1/3$
     curve would be vertical. Values of the termination
     condition in the range $0.1 \lesssim \epsilon \lesssim 1$
     give rise to variations that are narrower than the widths
     of the curves.  The light purple regions are below
     the electroweak scale $T_{\rm EW}\sim 100~{\rm GeV}$. The
     dark purple regions, below $10~{\rm MeV}$, would ruin the
     predictions of big bang nucleosynthesis (BBN).  Temperatures 
     above the intersection point are unphysical as they
     correspond to $N_{\rm re}<0$.  
     The gray shaded triangles indicate the parameter space allowed if $0<w_{\mathrm{re}}
     <0.25$.   The light yellow band indicates the $1\sigma$
     range in $n_s-1=-0.0397\pm 0.0073$ from
     Planck~\cite{Ade:2013uln}, and the dark yellow band assumes
     a projected uncertainty of
     $10^{-3}$~\cite{1404.1065} for $n_s-1$ as expected
     from future experiments (assuming the central value remains
     unchanged).  The top panels plot the number $N_k$ of
     $e$-folds of inflation as a function of $n_s$.   The
     vertical dashed red lines demarcate the allowed range of $n_s$, inferred from the lower panel,
     and the horizontal dashed red lines in the upper panels
     indicate the allowed range of values of $N_k$.}
\label{fig:natural}
\end{figure*}

\section{Inflaton potentials}

We now discuss the two classes of inflation models that we
consider in this work.

\subsection{Natural Inflation}

This model, first proposed in Ref.~\cite{Adams:1992bn},
appears when a global $U(1)$ symmetry is spontaneously broken.
The inflaton is then the pseudo-Nambu-Goldstone boson.
The shift symmetry protects the flatness of the potential.
The inflaton potentials we consider are,
\begin{equation}
     V(\phi) = \dfrac{2 \Lambda^4}{2^m} \left(
     1+\cos\phi/f\right)^{m},
\end{equation}
where the energy density $\Lambda^4$ and decay constant $f$ are
the parameters of the model.  We generalize the usual
natural-inflation potential, which has $m=1$, to other values of
$m$ to broaden slightly the class of models we consider.
The slow-roll parameters for this model are
\begin{equation}
     \e = m^2 \dfrac{e^{-x}}{2f^2(1-e^{-x})+m}, \qquad
     \mbox{where} \quad x=\dfrac{mN_k}{f^2},
\end{equation}
and
\begin{equation}
     \eta = \eta_V-\e = \dfrac{-m}{2f^2}\,
     \dfrac{2f^2(1-me^{-x})+m}{2f^2(1-e^{-x})+m}.
\end{equation}
These lead to the observables $r$ and $n_s-1$, which are
\begin{equation}
     r = 8 m^2 \dfrac{e^{-x}}{2f^2(1-e^{-x})+m},
\end{equation}
and
\begin{equation}
     n_s-1 =  -\dfrac m {f^2} - \dfrac{2m(m+1)\, e^{-x}}{2f^2(1-e^{-x})+m}.
\end{equation}
We will also need to calculate the number $N_k$ of $e$-folds that happen after a
mode with wavenumber $k$ exits the horizon, which is found to be
\begin{equation}
     N_k = \dfrac{f^2}{m} \log\left[ \dfrac 1
     {1+m/(2f^2)}\dfrac{(n_s-1)f^2-m^2}{(n_s-1)f^2+m} \right ].
\end{equation}

Even though the model has two parameters ($\Lambda$ and $f$) only one of them is free, 
since they are related through the amplitude of the scalar power spectrum. From the value 
of the potential $V_k$ at horizon exit we find $\La$ to be,
\begin{equation}
     \La = \left ( \dfrac 3 4 \pi^2 r A_s \left[\dfrac {2 f^2+n}
     {2f^2(1-e^{-m N_k/f^2}) + m}\right]^m\right )^{1/4}.
\end{equation}

In the $f\to\infty$ limit these potentials behave like pure power laws; i.e.,
\begin{equation}
V(\phi)\sim M^{4-2m}\phi^{2m} \quad \mbox{when} \quad f\to\infty,
\end{equation}
where $M$ is an energy scale that plays the role of $\Lambda$ and is also fixed.

\subsection{Higgs-like Inflation}

The potentials we consider for Higgs-like inflation are,
\begin{equation}
     V(\phi) = \Lambda^4 \left[1-(\phi/\mu)^2\right]^n,
\end{equation}
with slow-roll parameters,
\begin{equation}
     \e = \frac{2n^2 y}{\mu^2 (1-y)^2},
\end{equation}
and
\begin{equation}
     \eta= \eta_V-\e = \frac{2n[-1+(n-1)y]}{\mu^2 (1-y)^2}.
\end{equation}
The variable $y$ is defined as,
\begin{equation}
     y(\mu)\equiv\phi_0^2/\mu^2=-W\left(-g(\mu )\exp\left[{-g(\mu
     )-\frac{8 N_k}{\mu ^2}}\right]\right),
\end{equation}
where $W(z)$ is the Lambert $W$ function, and 
\begin{equation}
     g(\mu) \equiv \left(\phi_{\mathrm{end}}/\mu\right)^2= 1 + \dfrac{n^2}{\mu^2} -
     \dfrac{\sqrt{n^4+2\mu^2n^2}}{\mu^2}<1. 
\end{equation}
Again, we generalize the usual case ($n=2$) to explore a broader
class of models. In the general case the tensor-to-scalar ratio and scalar spectral index are,
\begin{equation}
     r = \frac{16n^2 y}{\mu^2 (1-y)^2},
\end{equation}
and
\begin{equation}
     n_s-1 =  -\dfrac {4n} {f^2} \dfrac{[1+(n+1)y]}{(1-y)^2}.
\end{equation}
We will again need the number,
\begin{equation}
     N_k =\frac{\mu ^2}{4n} \left[- \log
     \left(\frac{y}{g}\right)+y-g\right],
\end{equation}
 of $e$-folds of inflation, and once again we can express the amplitude $\La$ of the potential in terms of
the scalar power-spectrum amplitude $A_s$ and the decay constant $\mu$,
\begin{equation}
     \La = \left [ \dfrac 3 2 \pi^2 r A_s
     \left(1-y\right)^{-n}\right ]^{1/4}.
\end{equation}

This model also behaves as a power law in the $\mu\to\infty$ limit, the exponent being in this case $n$,
\begin{equation}
V(\phi)\sim M^{4-n}\phi^{n} \quad \mbox{when} \quad \mu\to\infty.
\end{equation}

\section{Results}
\begin{figure*}[htbp]
\centering
\hspace{-0.5cm}
\includegraphics[width=0.3\linewidth]{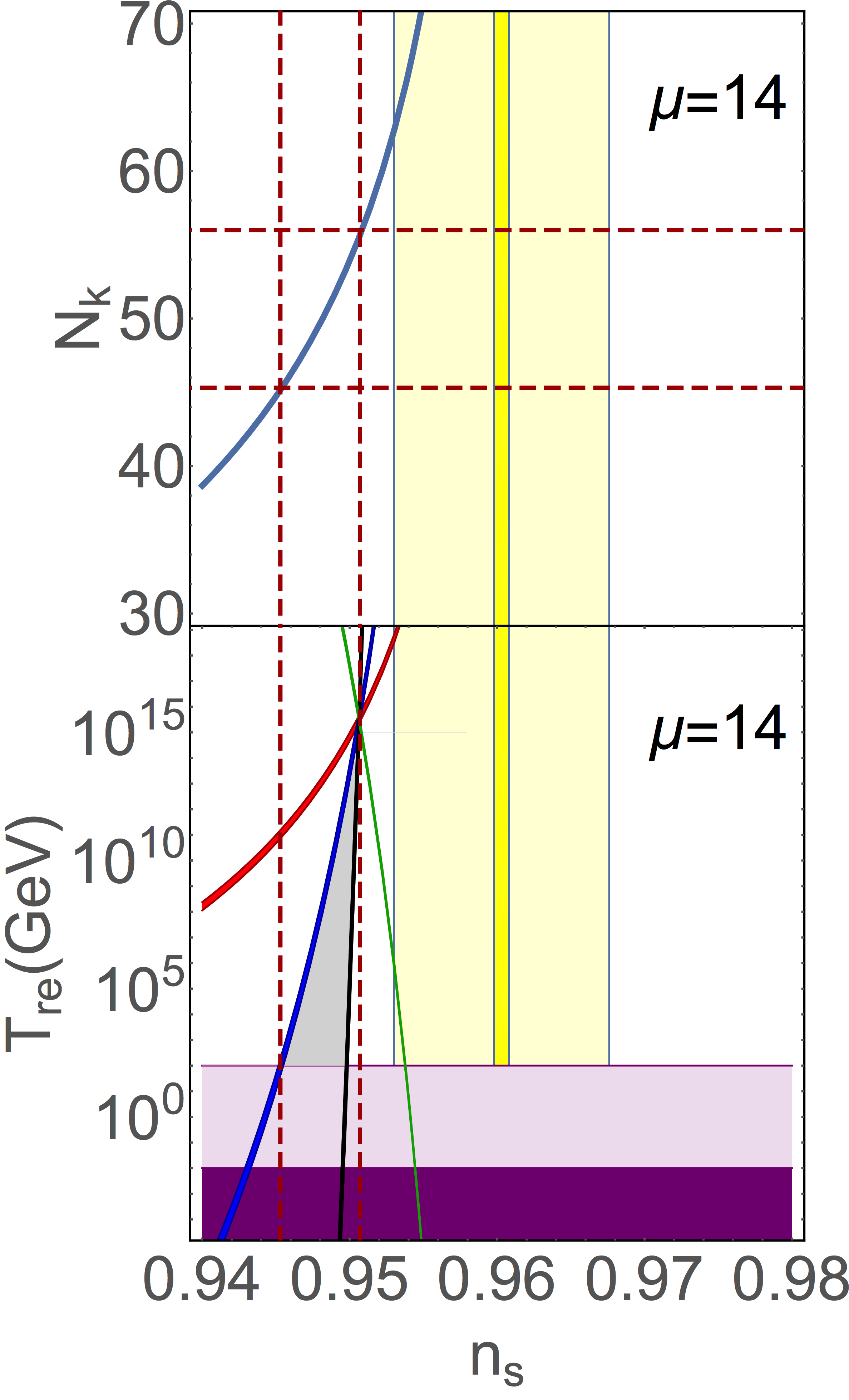}
\hspace{0.5cm}
\includegraphics[width=0.3\linewidth]{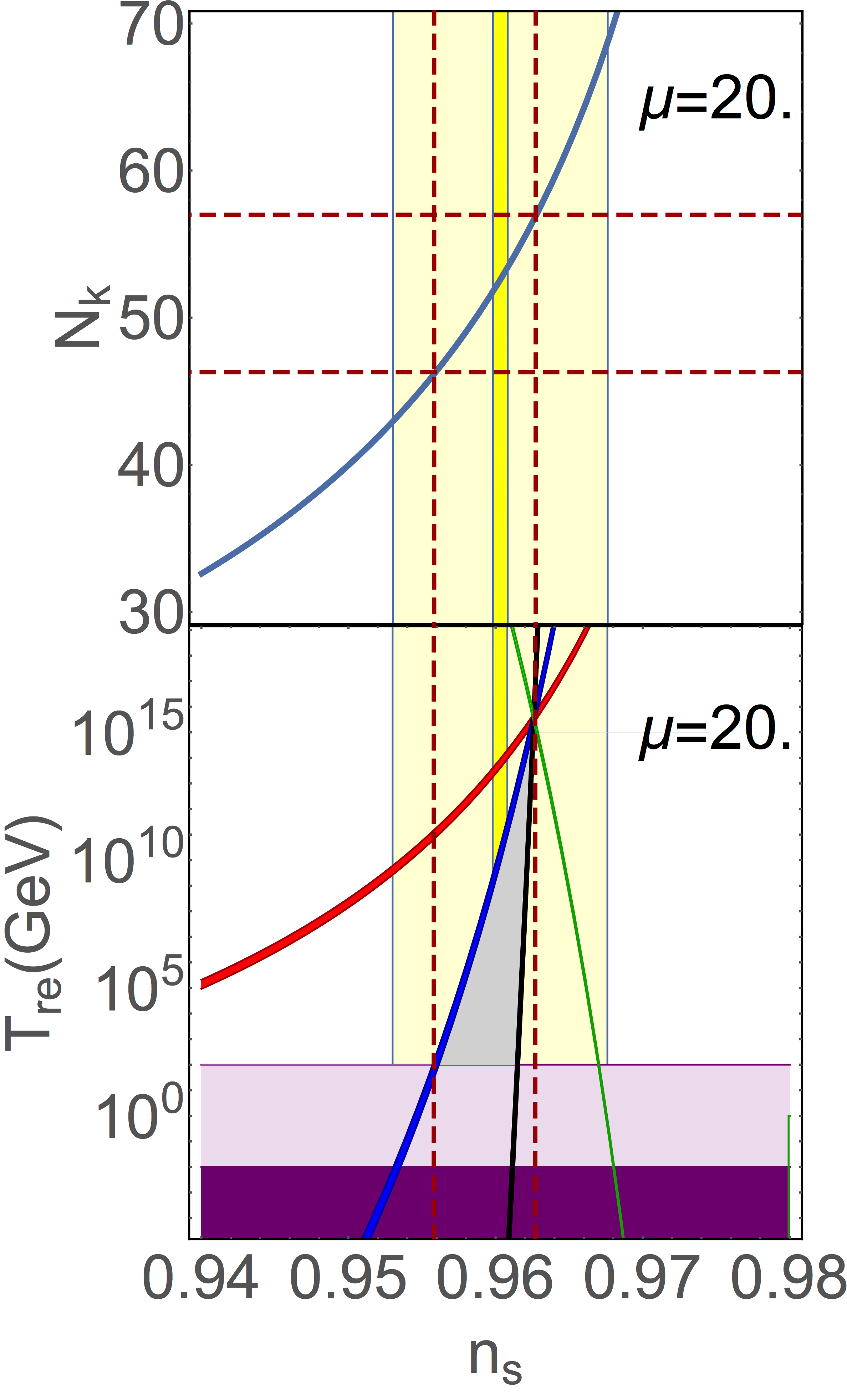}
\hspace{0.5cm}
\includegraphics[width=0.3\linewidth]{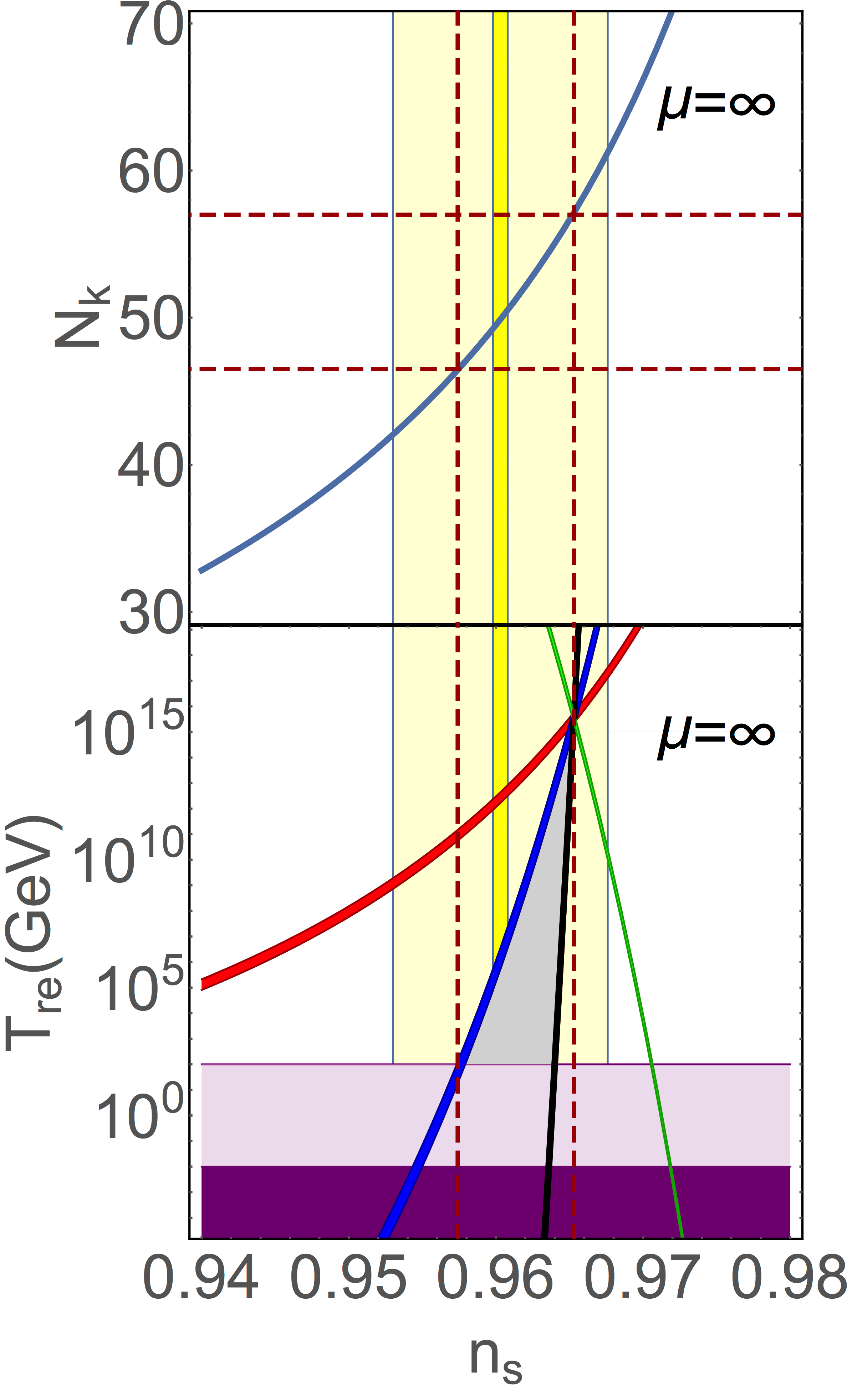}
\caption{Same as Fig.~\protect\ref{fig:natural} but for Higgs-like
     inflation with parameter values $\mu=14\, \M$, 20 $\M$, and $\infty$.}
\label{fig:higgs}
\end{figure*} 

The results of the calculation are shown for usual natural inflation
in Fig.~\ref{fig:natural} and for usual Higgs-like inflation in
Fig.~\ref{fig:higgs}. The reheat temperature $T_{\rm re}$ determined 
by matching the number of $e$-folds during and after inflation is shown 
in the lower panels of each Figure.  We show
results for four different reheating effective equation-of-state
parameters $w_{\mathrm{re}}$.  The value $w_{\mathrm{re}}=-1/3$ indicates
the smallest possible value of $w_{\mathrm{re}}$ required for
inflation to end.  The value $w_{\mathrm{re}}=1$ provides the most
conservative upper limit which comes simply from causality.  The
values $w_{\mathrm{re}}=0$ and $w_{\mathrm{re}}=0.25$ bracket the range of
values of $w_{\mathrm{re}}$ in detailed models of reheating.  The
curves for all values of $w_{\mathrm{re}}$ intersect at the point
where reheating is instantaneous, and the $w_{\mathrm{re}}=1/3$ curve
would be vertical and intersect this point.  The gray shaded
triangles indicate the region allowed if the reheating
equation-of-state parameter lies in the range $0<w_{\mathrm{re}}
<0.25$.   

The top panels of Figs.~\ref{fig:natural} and \ref{fig:higgs}
plot the number $N_k$ of $e$-folds during inflation for each
model and value of $f$ (for natural inflation) or $\mu$ (for
Higgs-like inflation).  It can be seen, in particular, that the
limit to the allowable range of values of $n_s$ imposed by
reheating considerations thus restricts the allowed range of
values of $N_k$.  The range of values of $N_k$ is generally 
smaller than the range $N_k\simeq46-60$ often assumed, being replaced (at our pivot scale $k=0.05$ Mpc$^{-1}$)
 by $N_k\simeq47-57$ for the large $f,\, \mu$ limit, and slightly smaller values for
 lower $f,\, \mu$.

It is also important to note that the tightness of the
constraint to the $n_s$ parameter space for fixed $f$ (for
natural inflation) or $\mu$ (for Higgs-like inflation) is
determined not by the precision of current measurements, but by
the self consistency of the inflationary-plus-reheating model. For the $m^2\phi^2$
case the new range of possible $n_s$ for  inflation is (0.958,0.965).

\begin{figure*}[htbp!]
\centering
\hspace{-0.5cm}
  \includegraphics[width=0.47\linewidth]{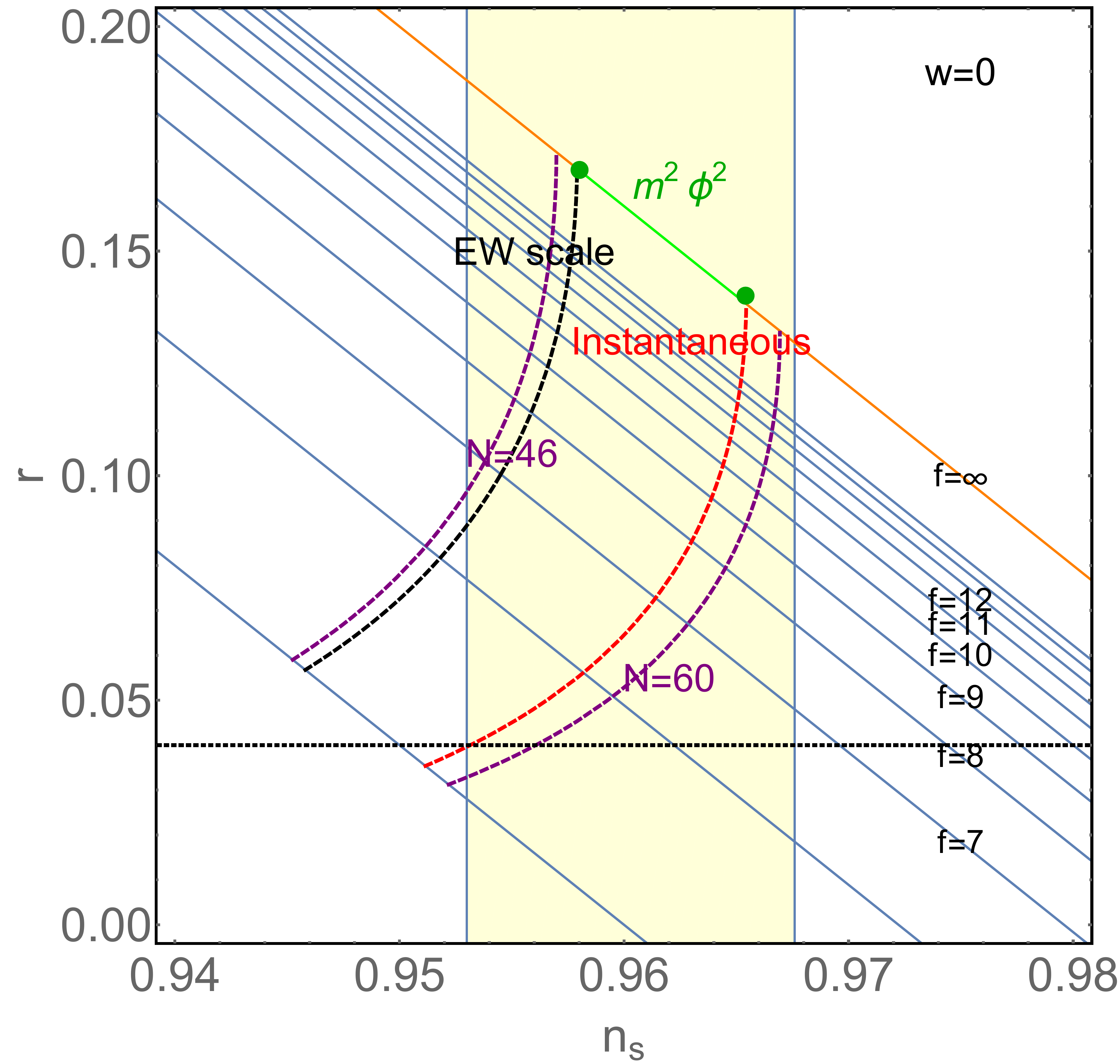}
\hspace{0.5cm}
  \includegraphics[width=0.47\linewidth]{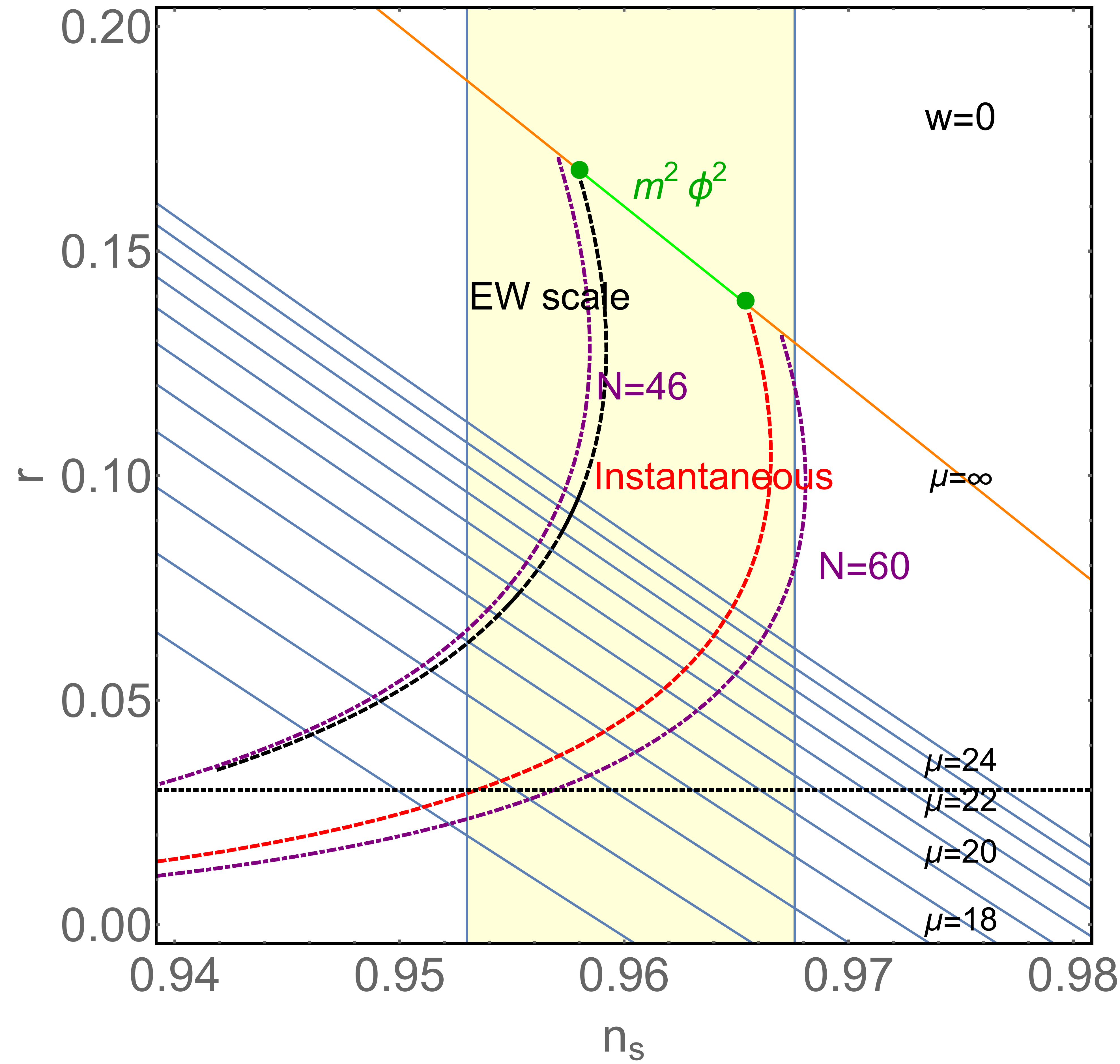}
  \caption{The $n_s$-$r$  parameter space for (left) natural
  inflation and (right) Higgs-like inflation.  Curves that
  indicate instantaneous reheating (red) and reheating at the
  electroweak scale (black) are shown as well as curves that show $N_k=46$
  and $N_k=60$ $e$-folds of reheating (purple).  Diagonal blue lines indicate different 
  values of the decay constants $f$ or $\mu$, where the orange line is the power-law limit.
  The horizontal dotted lines indicate the
  smallest tensor-to-scalar ratio $r$ consistent
  with the $1\sigma$ range of values of the scalar spectral
  index $n_s$, obtained by restricting the reheating
  equation-of-state parameter to physically plausible values, which are higher by about 25$\%$
  than those obtained by simply taking a range $N_k=46-60$ for the number of
  $e$-folds of reheating.}
\label{fig:rns}
\end{figure*}

We also show results in Fig.~\ref{fig:rns} as plots of the
$r$-$n_s$ parameter space for natural inflation and for
Higgs-like inflation.  It is seen here that even after
considering the complete range of values of $f$ (for natural
inflation) or $\mu$ (for Higgs-like inflation), the parameter
space allowed by restricting the reheating equation-of-state
parameter to physically plausible values is more constrained than
that assumed simply taking a range $N_k=46-60$ for the number of
$e$-folds of inflation.  In particular, we see that the smallest
tensor-to-scalar ratio $r$ allowed by the current $1\sigma$
range of values for $n_s$ is a bit larger with our approach than
that obtained with the less restrictive analysis. The black (dashed) curves correspond to 
the maximum reheating possible with equation-of-state parameter $w_{\mathrm{re}}=0$. Increasing 
the value of $w_{\mathrm{re}}$ would only shift
the black curves to the right.

\section{Conclusions}

We have explored a new technique to find constraints to inflationary models
by studying their reheating period. Instead of focusing on the physics of the 
reheating phase itself, or assuming an overly ample parameter space by constraining
the number of $e$-folds of inflation, we characterize the whole reheating era by a single
equation-of-state parameter $w_{\mathrm{re}}$, that we constrain to have physically reasonable values. 
This leads to more precise constraints to the inflationary 
observables. 

We have applied this formalism to two families of potentials (natural inflation and
Higgs-like inflation), finding better lower bounds for the tensor-to-scalar ratio $r$, as can be seen
in Table \ref{tab} (where the usual $m=1$, $n=2$ potentials are in bold face). 
It is important to notice that these results are robust to changes in the equation-of-state
parameter as long as it is kept under $w_{\mathrm{re}}=$1/3, as suggested by the literature on reheating.

The results derived for the potentials studied also apply, taking the 
limiting cases $f$ or $\mu \to \infty$, to power-law models and, as we show in
Figure \ref{fig:rns}, the allowed region for the power-law case (green line) is more constrained
using our method than with the usual analysis in which the range for the numbers of $e$-folds is fixed. 
For comparison, the right-hand plots in Figures \ref{fig:natural} and \ref{fig:higgs} correspond to the 
plot made on \cite{Dai:2014jja} for $m^2\phi^2$ potential, showing in the upper panel $N_k$ instead of $N_{\rm re}$.

\begin{center}
\begin{table}[h]
    \begin{tabular}{| l | l | l | }
    \hline
    Model & $r_{\mathrm min}$ old & $r_{\mathrm min}$ new \\ \hline
    Higgs $n=1$ & 0.020 & 0.025  \\ 
  \bfseries{Higgs $\mathbf n =$ 2} & \bfseries 0.024 & \bfseries 0.030 \\ 
    Higgs $n=3$ & 0.035 & 0.050  \\ 
    Higgs $n=4$ & 0.055 & 0.070  \\ 
      \bfseries{Natural $\mathbf m =$ 1}  &\bfseries 0.033 &\bfseries 0.040  \\ 
    Natural $m=3/2$ & 0.055 & 0.070  \\ 
    Natural $m=2$ & 0.10 & not allowed  \\  
     $m^2\phi^2$ & 0.13 & 0.14  \\     
	\hline
    \end{tabular}
\caption{Minimum value of the tensor-to-scalar ratio $r$ at the pivot scale $k=0.05$ Mpc$^{-1}$
allowed by reheating considerations and the Planck $1\sigma$ range of values 
of the scalar spectral index $n_s$ for each of the models studied.
In the central column we show the minimum $r$ from the usual analysis in which a range of $N_k$ is allowed,
and in the right column the new minimum obtained by constraining the reheating equation-of-state.}
\label{tab}
\end{table}
\end{center}

The most interesting feature of this technique is its general validity.  It was considered for 
power-law potentials in Refs.~\cite{Dai:2014jja,Creminelli:2014fca}, and we have generalized 
here to natural and Higgs-like potentials.  Still, the approach can be similarly applied to any 
single-field inflation model and will generically lead to slightly more restrictive bounds to the 
inflationary parameter space, including the range of values of the tensor-to-scalar ratio $r$.  
As a result, upper bounds to $r$, for example, will generally be slightly more restrictive to 
inflationary models than they would otherwise be.

\smallskip
\begin{acknowledgments}
The authors wish to thank Liang Dai and Ely Kovetz for useful comments on a previous draft.
This work was supported by the John Templeton Foundation, the
Simons Foundation, NSF grant PHY-1214000, and NASA ATP grant
NNX15AB18G.
\end{acknowledgments}

\end{document}